\documentclass[submission,copyright,creativecommons]{eptcs}
\providecommand{\event}{FROM 2026} 

\usepackage{iftex}
\usepackage{listings}
\usepackage{amsmath}
\usepackage{amssymb}
\usepackage{graphicx}

\usepackage{amsthm}
\theoremstyle{definition}
\newtheorem{exmp}{Example}[section]

\ifpdf
  \usepackage[strings]{underscore}         
  \usepackage[T1]{fontenc}        
\else
  \usepackage{breakurl}           
\fi

\usepackage{titlesec}
\titlespacing{\paragraph}{%
  0pt}{
  0.5\baselineskip}{
  1em}

\providecommand{\keywords}[1]{%
  {\small\textbf{\textit{Keywords---}} #1}%
}

\title{Formalizing the Omega Test in Dafny\\
\large Short Paper}
\author{Ariadna Brănici-Faraon \qquad\qquad Ștefan Ciobâcă \qquad\qquad Diana-Elena Gratie {
  
  \footnote{Authors in alphabetical order.}
}
\institute{Alexandru Ioan Cuza University of Iași, Romania}
\email{branici.ariadna6@gmail.com \qquad\qquad stefan.ciobaca@uaic.ro \qquad diana.gratie@uaic.ro }}

\def\titlerunning{Formalizing the Omega Test in Dafny}
\def\authorrunning{A. Brănici, Ș. Ciobâcă, D.E. Gratie}

\begin{document}
\maketitle

\begin{abstract}
    We present a formalization in Dafny of the Omega Test, an algorithm used to decide the satisfiability of a system of inequalities. The implementation defines executable representations for rational numbers, linear expressions, inequalities, equalities, divisibility constraints, and systems of constraints, together with their semantic interpretation through valuations. We fully specify and verify the implementation in Dafny. We describe the lessons learned and how the formalization process led to new insights into the algorithm.
\end{abstract}


\section{Introduction}
\paragraph{}Deciding if a system of inequalities has an integer solution is a classical problem that frequently appears in program analysis, compiler optimizations and SMT solvers. Modern SMT solvers, such as Z3~\cite{z3} and cvc5~\cite{cvc5}, include procedures to reason about linear integer arithmetic. A standard algorithm for this is the \emph{Omega Test}~\cite{pugh_omega, dillig_omega}, which eliminates variables one at a time using projections based on \emph{Fourier--Motzkin}~\cite{dantzig1973fourier, williams1976fourier} elimination. However, solving for integers becomes more difficult, since a non-empty rational interval might not contain any integer value. 

This paper presents a Dafny formalization of the main components of the \emph{Omega Test} for systems of linear inequalities, building on the BSc. thesis of the first author~\cite{branici_thesis}. The implementation presented contains executable representations for rational coefficients, linear expressions, and constraint systems, along with their semantic interpretations. Using these definitions, the algorithm's three key projections are implemented and verified: the Real Shadow, Dark Shadow, and Grey Shadow. 

The primary focus of the paper is the Dafny verification structure that sustains these transformations, along with some mathematical proof specifications. The Real Shadow is an over-approximation, used for deciding unsatisfiability of the system. The Dark Shadow serves as an under-approximation and ensures that valid solutions can be lifted back to the original variables and prove satisfiability. Any boundary cases that exist in the case of the Dark Shadow failure are systematically addressed by the Grey Shadow branches, which are executed by using additional equality constraints. 
There already exist formal and semi-automated treatments of Presburger arithmetic in proof assistants. For instance, Rocq previously provided the \emph{omega} tactic\cite{rocqOmega}, which was later deprecated in favor of the more general \emph{lia} tactic, a complete solver for Presburger arithmetic. However, the purpose of this work is not to compete with existing SMT solvers or Presburger arithmetic solvers in terms of performance or generality, but to study the \emph{Omega Test} as an implementation of an elimination-based procedure that utilizes explicit contracts, invariants, and ghost predicates with the aim of bridging the gap between the executable code and its underlying mathematical meaning. 

Section~\ref{sec:related} presents related implementations of the \emph{Omega Test}, Section~\ref{sec:Omega} illustrates an example of the \emph{Omega Test} on a system with two variables and three inequalities, Section~\ref{sec:FormalRep} details the formal representations, Section~\ref{sec:verifiedElimination} describes the verified elimination procedure, Section~\ref{sec:termination} explains the termination proof for the recursive method used in this implementation and Section~\ref{sec:Limitations} discusses limitations and outlines potential future work.

The Dafny implementation of the \emph{Omega Test} that we describe is available at \begin{center}\url{https://github.com/ariadna-b/FourierMotzkin}.\end{center}

\section{Related Work}\label{sec:related}
\paragraph{}The Omega Test and related procedures for Presburger arithmetic have been integrated into several proof assistants. Rocq previously provided the \emph{omega}~\cite{rocqOmega} tactic, whose decision procedure implemented only a part of Pugh's Omega Test, without the final steps of the Grey Shadow. The tactic was later deprecated in favor of \emph{lia}~\cite{rocqLia}, which provides a complete procedure for Presburger arithmetic.

Lean~\cite{leanOmega} also provides an \emph{omega} tactic for linear arithmetic over integers. Its implementation follows Pugh's Omega algorithm, but currently does not implement the Dark and Grey Shadows and is not a complete decision procedure.

Isabelle/HOL provides automated support for Presburger arithmetic through methods such as \emph{presburger}~\cite{isabellePresburger}. These procedures eliminate integer variables using periodicity and divisibility properties rather than the shadow projections of the \emph{Omega Test}.

The aim of the implementation presented in this paper is not to provide a general Presburger arithmetic solver, but to make the elimination steps of the \emph{Omega Test} specific in Dafny and to verify the semantic correctness.

\begin{figure}[ht!]
    \centering
    \includegraphics[width=1\textwidth]{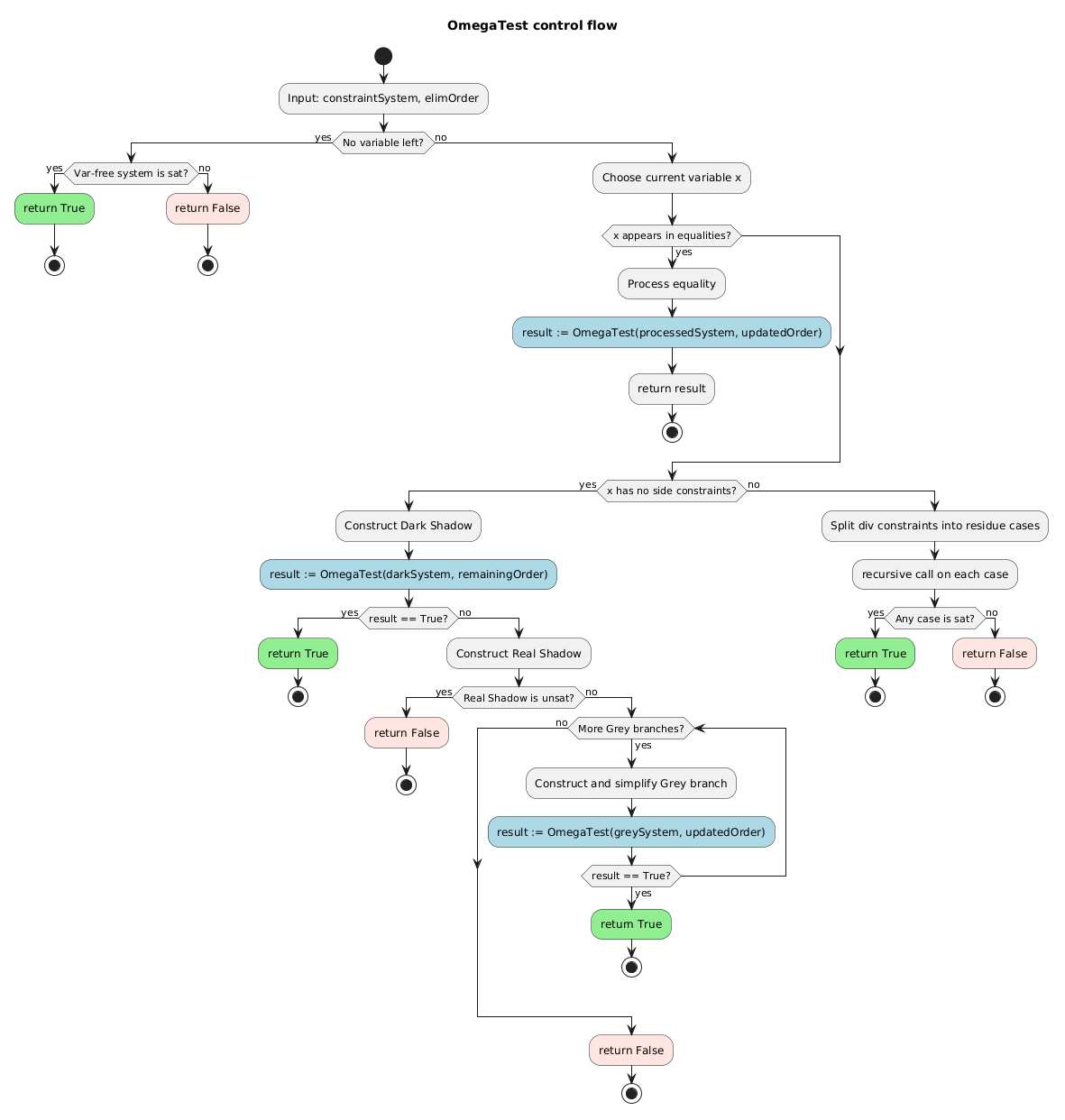}
    \caption{The Workflow of the Omega Test}
    \label{fig:omega-flow}
\end{figure}
\section{The Omega Test}\label{sec:Omega}
\paragraph{}The \emph{Omega Test} algorithm solves the problem of finding integer solutions to a system of linear equalities and inequalities by using three projections: \emph{Real Shadow}, \emph{Dark Shadow} and \emph{Grey Shadow}. The \emph{Real Shadow} over-approximates the solution, so if the projection has no solution, then neither does the original system. The \emph{Dark Shadow} under-approximates the solutions: if we find one in the projection, then it is guaranteed the original system also has one. If the \emph{Real Shadow} finds a solution, but the \emph{Dark Shadow} does not, the algorithm still checks the regions between the two approximations. Therefore, the \emph{Omega Test} combines a super-approximation, an under-approximation and a finite exploration of edge cases to decide the satisfiability of a system over integer numbers.

We use $L$ and $U$ for the expressions defining the lower and upper bounds. Before the shadow projections are constructed, the constraints are normalized and brought to a form more suitable for variable elimination. After preprocessing, consider a lower and an upper bound for the variable $x$ of the form $L \leq ax, \qquad bx \leq U, \qquad a,b>0$. The \emph{Real Shadow} eliminates $x$ by requiring $bL \leq aU$. The \emph{Dark Shadow} strengthens this condition to $aU-bL>ab-a-b$, guaranteeing the existence of an integer value for $x$. When the Real Shadow is satisfiable, but the Dark Shadow is not, the \emph{Grey Shadow} explores the remaining boundary cases through equations of the form $ax = L + i$, where $0 \leq i \leq \dfrac{ab-a-b}{b}$ for each lower bound $L\leq ax$ and upper bound $bx \leq U$.

Unlike typical presentations of the Omega Test ~\cite{kroening2008decision, dillig_omega}, where the \emph{Real Shadow} is constructed before the \emph{Dark Shadow}, our implementation constructs the \emph{Dark Shadow} first. This ordering was chosen to facilitate the proof of the satisfiability argument on the recursive \emph{Dark Shadow} call. Figure~\ref{fig:omega-flow} summarizes the interaction between the three shadows and highlights the recursive calls performed by the algorithm.

The input \texttt{constraintSystem} is the current system processed by the algorithm, containing not only inequalities, but also equalities and divisibility constraints. The sequence \texttt{elimOrder} stores the variables that still have to be eliminated. At each recursive step, the first variable of this sequence is selected as the current variable $x$. When $x$ is removed from the system, the recursive call uses \texttt{remainingOrder}, which represents the tail of the elimination order, without $x$.

The notation \texttt{updatedOrder} is used in the diagram for recursive calls where the order changes in a different way. For example, when a divisibility constraint containing $x$ is processed, the algorithm splits the problem into finitely many cases of the form $x = mq + r$, where $m$ is the product of the divisors in the current system, $r$ is one residue with $0 \leq r < m$ and $q$ is a fresh quotient variable. In this case, $x$ is replaced by $q$, so the recursive call uses an updated order of the form \texttt{[q] + \texttt{remainingOrder}}. For the grey branches, \texttt{updatedOrder} depends on the simplification step: if the grey equality can be solved for $x$ or if new divisibility constraints are generated.

More details about how the shadows interact with equality and divisibility constraints, and how different cases are handled in the implementation, are given in Section~\ref{sec:verifiedElimination}. 

In Example~\ref{exp:System} we show how the algorithm constructs each shadow and decides satisfiability for a given system: 
\begin{exmp}\label{exp:System}  
\[
\begin{aligned}
S &= \left\{
\begin{array}{rcl}
5y - x + 7 &\leq& 2x + 2y + 4,\\
4x + y + 1 &\leq& 2x + 3y + 3,\\
2x - y &\leq& 2x
\end{array}
\right\}
\xrightarrow{\text{Normalize}}
\left\{
\begin{array}{rcl}
3y-3x &\leq& -3,\\
2x-2y &\leq& 2,\\
-y  &\leq& 0
\end{array}
\right\} \\[1em]
S &\xrightarrow{\text{Dark eliminate }x}
D_1 = \left\{
\begin{array}{rcl}
-y &\leq& 0,\\
0  &\leq& -2
\end{array}
\right\} 
\;\xrightarrow{\text{eliminate }y}
D_2 = \left\{ 0 \leq -2 \right\}
\quad \Rightarrow \quad \textbf{False} \\[1em]
S &\xrightarrow{\text{Real eliminate }x}
R_1 = \left\{
\begin{array}{rcl}
0  &\leq& 0,\\
-y &\leq& 0
\end{array}
\right\}
\;\xrightarrow{\text{eliminate }y}
R_2 = \left\{
0 \leq 0
\right\}
\quad \Rightarrow \quad \textbf{True} \\[1em]
S &\xrightarrow{\text{Grey Shadow}}
G_1 = \left\{
\begin{array}{rcl}
3y-3x &\leq& -3,\\
2x-2y &\leq& 2,\\
-y  &\leq& 0,\\
3x   &=& 3y+3
\end{array}
\right\}
\xrightarrow{\text{GCD simplification}}
G_2 = \left\{
\begin{array}{rcl}
3y-3x &\leq& -3,\\
2x-2y &\leq& 2,\\
-y  &\leq& 0,\\
x   &=& y+1
\end{array}
\right\} \\[0.5em]
&\xrightarrow{\text{substitution}}
G_3 = \left\{
\begin{array}{rcl}
3y-3(y+1) &\leq& -3,\\
2(y+1)-2y &\leq& 2,\\
-y  &\leq& 0
\end{array}
\right\}.
\end{aligned}
\]
\end{exmp}
\paragraph{}The first operation is the normalization step. Each inequality is rewritten into the form $a_1x_1+\cdots + a_nx_n \leq c$, with all variable terms on the left and the constant term on the right. Here, for example, $5y - x + 7 \leq 2x + 2y + 4$ becomes $3y-3x \leq -3$.

The second transition constructs the \emph{Dark Shadow} with respect to $x$, using the first two inequalities as a lower ($L\leq  ax$) and an upper ($bx\leq U$) bound and using the formula $ab - a - b + 1 \leq aU - bL$, equivalent to $aU-bL>ab-a-b$, to obtain $2 \leq 0$, which is then normalized to $0 \leq -2$. Since the last normalized inequality does not contain $x$, in this step it just gets stored in the new system with no processing.

The same elimination is done for $y$ and the variable-free projection that is obtained is $0 \leq -2$. Since the constraint is false, the dark projection is unsatisfiable.

The algorithm then constructs the \emph{Real Shadow} by also eliminating first $x$, starting from the previously normalized system, this time using the formula $bL \leq aU$, equivalent to $\dfrac{L}{a} \leq \dfrac{U}{b}$. The first two inequalities result in a weaker constraint $0 \leq 0$ and the third, $-y\leq 0$, is simply stored in the system.

The new two inequalities are then processed again to eliminate $y$, obtaining $0 \leq 0$. Since the dark projection is false and the real one is true, we have to explore the \emph{Grey Shadow}.

For the lower bound $3y+3 \leq 3x$, the grey branch adds an equation of the form $3x = 3y + 3 +i $. For $i=0$, this becomes $3x = 3y+3$. The grey equalities are checked by a GCD simplification before the recursive call. In this case, the equality becomes $x = y + 1$ and because here the coefficient of the variable $x$ is $1$, the algorithm substitutes $x$ in the system. The \emph{Omega Test} is called again on the new system and, since it has a solution (for example $x=1$ and $y=0$), the grey branch returns \verb|True|, and the original system is satisfiable over integers.

In other cases when the coefficient of $x$ is not $1$, substitution is not possible and the algorithm derives new divisibility constraints from the equation. More implementation details behind these transitions are described in Section~\ref{sec:verifiedElimination}.

\section{Formal Representation}\label{sec:FormalRep}
\paragraph{}In Dafny, executable code is accompanied by specifications such as preconditions, written using \texttt{requires}, postconditions, written using \texttt{ensures}, and loop invariants, written using \texttt{invariant}. Dafny also supports ghost predicates and auxiliary lemmas used only for verification. During verification, Dafny generates proof obligations from these annotations and discharges them using an SMT solver. Termination of loops and recursive methods can additionally be specified using \texttt{decreases} clauses. In this work, the verification of the Omega Test relies on precise formal representations of linear expressions, inequalities and systems of constraints, assigning them a semantic interpretation through valuations, implementing the elimination transformations, and proving that these transformations preserve the required satisfiability properties.

Due to the fact that Dafny's \texttt{real} type is non-executable and therefore unsuitable for the computational part of the implementation~\cite{dafny_ref}, rational numbers are represented using a custom datatype called \texttt{Rational}, together with necessary functions that perform arithmetic operations over objects of this type. An object of this type contains a numerator, a strictly positive denominator and a \texttt{real} value used only for specifications:
\begin{lstlisting}[frame=tb, belowskip=0pt]
type pos = n: int | n > 0 witness 1 
datatype Rational = Rational(num: int, den: pos, val: real) 
ghost predicate validRational(r: Rational) 
{ 
    r.val == r.num as real / r.den as real 
} 
\end{lstlisting}
\paragraph{}
The type \texttt{pos} ensures that the denominator cannot be zero or negative. The ghost predicate \texttt{validRational} states the connection between the two concrete fields (\texttt{num}, \texttt{den}) and the mathematical value represented by their fraction (\texttt{val}). This predicate must be proven to be true in all operations performed on the objects of type \texttt{Rational} and in other declarations that use those operations. For example, the multiplication function guarantees this relation:
\begin{lstlisting}[frame=t, belowskip=0pt]
function multiply(rat1 : Rational, rat2 : Rational) : Rational
    requires validRational(rat1)
    requires validRational(rat2)
    ensures validRational(multiply(rat1, rat2))
    ensures multiply(rat1, rat2).val == rat1.val * rat2.val
{
    var den' := rat1.den * rat2.den;
    var num' := rat1.num * rat2.num;
    Rational(num', den', num' as real / den' as real)
}
\end{lstlisting}
\paragraph{}The same module (\texttt{rational}) also defines the conversion from \texttt{int} numbers to \texttt{Rational}, the recursive sum of a sequence, the semantic equality of two \texttt{Rational} values and the ceiling of a \texttt{real} (by applying $\lceil r \rceil = -\lfloor -r \rfloor$). The equality of two \texttt{Rational} values is not verified by direct comparison of numerators and denominators, because more representations for the same values could exist: \(\dfrac{1}{2}\) and \(\dfrac{2}{4}\). Instead, the predicate \texttt{RationalEq} compares the mathematical values of the two objects by subtracting them: \texttt{subtract(x,y).val == 0.0}. There are lemmas that prove properties such as reflexivity, symmetry, transitivity, commutativity, associativity and distributivity, needed to verify future operations on linear expressions.

Other important modules necessary for the formal representation of linear algebra used in the project are \texttt{dataTypes} and \texttt{evaluating}, which define the main structures used in the formalization of systems of inequalities. The atomic unit is defined by a \texttt{Term}, which associates a \texttt{Rational} coefficient to a variable name. Terms are grouped in linear expressions, which are sequences of \texttt{Terms} and a \texttt{Rational} constant. An inequality is formed by two linear expressions, mostly represented as $L \leq U$, other forms are used only in bridging proofs. There also exist data types for representing equalities and divisibility constraints, named below.
\begin{lstlisting}[frame=tb , belowskip=0pt]
type Valuation = string -> int
Term(name: string, coeff: Rational) 
LinearExpr(terms: seq<Term>, constant: Rational)
Inequality(left: LinearExpr, right: LinearExpr)
EqualityConstraint(left: LinearExpr, right: LinearExpr)
DivisibilityConstraint(divisor: int, expr: LinearExpr)
\end{lstlisting}
\paragraph{}Evaluating an \texttt{Inequality} is done gradually, starting with \texttt{EvalTerm}, which just multiplies the coefficient of the term with the integer value associated to the given valuation of the variable. A \texttt{LinearExpr} is evaluated by using a recursive function that sums a sequence of terms and adding the constant to the result. Finally, \texttt{EvalInequality} is a boolean function that verifies if the difference between the left side of an inequality and its right side is non-positive. Using sequences instead of a function $variable \rightarrow coefficient$ allows executable traversal of terms and incremental construction of linear expressions during normalization and variable elimination. The code snippet below exemplifies how some of the functions in the module \texttt{evaluating} are defined:
\begin{lstlisting}[frame=tb, belowskip=0pt]
// EvalTerm:
multiply(t.coeff, IntToRational(val(t.name)))
// EvalExpr:
add(SumTerms(e.terms, val), e.constant)
// EvalInequality:
var diff := subtract(EvalExpr(i.left, val), EvalExpr(i.right, val))
diff.val <= 0.0
\end{lstlisting}
\paragraph{}We use a variety of predicates for identifying forms of the inequality system, such as \texttt{ValidTerms} (all the terms in a sequence have valid \texttt{Rational} coefficients), \texttt{UniqueTerms} (variable names do not repeat in a sequence of terms), \texttt{ValidInequality} (both linear expressions forming it have valid terms and valid constants),\\ \texttt{NormalizedInequality} (of the form $a_1x_1 + ... + a_nx_n \leq c$ with no variable duplicates).

Because the problem of deciding the satisfiability of a system of constraints implies processing the system in different ways, there are several ways of representing the systems. The simpler representation is \texttt{IneqSystem}, which is just a sequence of inequalities and is used by the Real, Dark and Grey Shadow constructions. The main recursive method \texttt{OmegaTest} works over a more general \texttt{ConstraintSystem}, which contains three parts: inequalities, equality constraints and divisibility constraints. This  is needed because the \emph{Grey Shadow} may introduce equalities of the form $ax = L + i$, simplify them to the form $a'x = E$, and generate divisibility conditions from them, such as $a' \mid E$.

For the elimination steps on inequalities, there exists the type \texttt{PartitionedSystem}, which contains three sequences of inequalities, separated depending on the sign of the coefficient of a certain variable (positive coefficient, negative coefficient and an inequality where the variable does not appear or has the coefficient equal to 0). Also, \texttt{PairDark} and \texttt{DarkTraceSystem} express the source pair of inequalities used by the \texttt{Dark Shadow} to generate a projection. The definitions of those types are presented below.
\begin{lstlisting}[frame=tb, belowskip=0pt]
    datatype IneqSystem = IneqSystem(ineqs: seq<Inequality>)
    datatype ConstraintSystem = ConstraintSystem(
        ineqs: seq<Inequality>,
        equalities: seq<EqualityConstraint>,
        divisibilities: seq<DivisibilityConstraint>)
    datatype PartitionedSystem = PartitionedSystem(
        pos: seq<Inequality>,  
        neg: seq<Inequality>,
        zero: seq<Inequality>)

    datatype PairDark<A, B, C> = PairDark(a: A, b: B, c: C)

    datatype DarkTraceSystem = DarkTraceSystem(
        base: seq<Inequality>,
        pairs: seq<PairDark<Inequality, Inequality, Inequality>>)
\end{lstlisting}
\paragraph{}Equivalence between objects is defined semantically by their behavior for all possible evaluations of the variables and not by the structural equality of the representations. Two sequences of \texttt{Terms} are not considered equivalent only in the case they have the same exact variables, with the same coefficients, but if for all possible valuations, they have the same sum: $\forall val: SumTerms(T_1, val) == SumTerms(T2, val)$. Similarly, two linear expressions are equivalent if their evaluations are the same for any variable assignment. For inequalities, the equivalence is decided if they have the same truth value under any valuation, even if they are represented in different syntactical forms. Finally, two systems are equivalent if they are satisfied by the same valuation: $\texttt{EquivalentSystems}(S_1,S_2) \iff \forall val: \texttt{SystemHolds}(S_1,val) == \texttt{SystemHolds}(S_2,val)$.

Defining equivalence like this is essential for verifying the transformations performed over the objects, because normalization, reordering or grouping terms, moving them from one side of the inequality to the other or deleting duplicate terms can modify the internal structure of an object, but not its mathematical meaning. 

\section{Verified Elimination}\label{sec:verifiedElimination}
\paragraph{}The Fourier--Motzkin elimination is a procedure of variable elimination in linear inequalities. The implementation follows and extends the presentation of the algorithm used in the Omega Test lecture notes by Dillig~\cite{dillig_omega}. For the first variable in the elimination order, the method \texttt{OmegaTest} first checks whether the variable occurs in an equality. If a unit equality of the form $x = L$ is found, the variable is substituted in the system and the recursive call continues with the tail of the elimination order. If the equality has a non-unit coefficient, the implementation first applies the GCD compatibility test: incompatible equalities immediately prove unsatisfiability, while compatible ones are rewritten into a form that removes the current variable and adds the new divisibility obligation.

For example, an equality such as $x = 2y + 3$ allows the algorithm to substitute $2y + 3$ for every occurrence of $x$, which can now be removed from the elimination order. If the equality is instead $2x = 4y + 6$, the GCD simplification will transform it into $x = 2y + 3$ and again substitute $x$ with $2y + 3$ in the system.

When the current variable does not occur in equalities or divisibility constraints, the method enters the shadow part of the Omega Test. It takes the chosen variable $x$ and classifies each inequality according to the sign of the coefficient of $x$ and memorizes it in an object of type \texttt{PartitionedSystem}. The \texttt{pos} inequalities determine the superior limit for $x$, the \texttt{neg} inequalities determine the inferior limit, and the \texttt{zero} are the ones in which $x$ does not appear (inequalities that do not depend on $x$). After isolating the variable, each inferior limit has to be compatible with the superior limit. Using the notation introduced in Section~\ref{sec:Omega}, for a lower bound $L \leq ax$ and an upper bound $bx \leq U$, the Real Shadow generates the constraint $bL \leq aU$. The projected system contains all inequalities that do not depend on $x$ (\texttt{zero}), together with a new deduced inequality for all pairs of inferior and superior limits of $x$. By checking the satisfiability of this new system, we can determine whether the original system has real solutions for $x$ or not. However, to check for an integer solution, the determined interval by an inferior limit and a superior limit might not be large enough to contain integer values. Therefore, the Fourier--Motzkin elimination only produced the \emph{Real Shadow} and needs to be completed with other projections used in the \emph{Omega Test}.

The implementation of the \emph{Real} projection is divided in more modules, each corresponding to a distinct stage of the elimination. The module \texttt{manipulateIneq} contains basic operations over expressions and inequalities, such as moving a constant from one side to the other, negating the terms or multiplying an inequality by a positive factor. Those transformations are followed by post-conditions that ensure the validity of the structure and the semantic equivalence. This makes possible modifying constraints without modifying their solution space.

Before elimination, each inequality in a system is normalized using definitions from the module \texttt{normalize}. The method \texttt{MoveAllTermsLeft} moves every term on the left side of the inequality by changing its sign and attributing it to a linear expression. After this procedure, the constant is also moved to the left side, the new constant calculated and moved back to the right side of the inequality. The function \texttt{Normalize} groups every appearance of the same variable and replaces the terms with the same variable name by a single one, whose coefficient is the sum of the previous coefficients: $5x, 2y, -1x,7z$ is transformed into $4x,2y,7z$. Finally, terms that resulted in a zero coefficient are being eliminated by \texttt{RemoveZeroTerms}. 
This way, an arbitrary constraint is transformed in an equivalent inequality of the form $a_1x_1 + ... + a_nx_n \leq c$ with no repeating variables, no null-coefficient term and a single constant on the right. Because it is important to keep track of all facts in the post-conditions, the final normalizing method ensures through predicates that the resulting system is valid, normalized, does not introduce new variables and is equivalent to the original system. Below, these transformations are exemplified on an inequality from Example~\ref{exp:System}. 
 
\[
\resizebox{\textwidth}{!}{$
\begin{aligned}
&5y - x + 7 \leq 2x + 2y + 4\text{   }
\xrightarrow{\;\texttt{NegateTermSeq}\;}
\text{   }
5y - x + 7 + (-2x) + (-2y) \leq 4 \text{   }
\xrightarrow{\;\texttt{MoveAllTermsLeft}\;}\text{   }
5y - x - 2x - 2y + 7 \leq 4\text{   } \\[0.4em]
&\xrightarrow{\;\texttt{MoveConstantOnRight}\;}\text{   }
5y - x - 2x - 2y \leq -3\text{   }
\xrightarrow{\;\texttt{Normalize}\;}\text{   }
(5-2)y + (-1-2)x \leq -3\text{   }
\xrightarrow{\;\texttt{RemoveZeroTerms}\;}\text{   }
3y - 3x \leq -3.
\end{aligned}
$}
\]

\paragraph{}To eliminate one variable $x$, the module \texttt{partition} first isolates that variable, depending on the sign of its coefficient, and uses a \texttt{PartitionedSystem} to store the resulting system. The methods \texttt{PartitionPosIneq} and \texttt{PartitionNegIneq} construct an equivalent inequality of the form $L \leq ax$ or $bx \leq U$ and the method calling them to form the system, \texttt{SplitSystem}, ensures facts such as: the \texttt{pos}, \texttt{neg}, \texttt{zero} fields contain the inequalities in their correct form; each inequality is equivalent to some inequality in the original system; the \texttt{ps.zero} inequalities are normalized and do not contain $x$. This is necessary for later stating that any solution that satisfies the original system also satisfies the partitioned one. 

After partitioning, each superior limit is combined with each inferior one. For a pair $L \leq ax $ and $bx \leq U$, the resulting inequality is $bL \leq aU$, which will be later normalized to $bL.terms - aU.terms\leq aU.const - bL.const$. The method \texttt{BuildProjectedPairs} covers the cartesian product of \emph{lower bounds $\times$ upper bounds} and adds in the new system a constraint for each pair, then the rest of inequalities from \texttt{zero}. To implement this in Dafny, the two while loops have strong invariants about how all generated inequalities are valid, do not contain \texttt{x} and are satisfied by any valuation that satisfies the original pair. The method forms the final projection for one variable, eliminating it from the system and finally stating:
\begin{lstlisting}[frame=tb, belowskip=0pt]
ensures forall val :: SystemHolds(sys, val) ==> SystemHolds(proj, val)
ensures forall val :: !SystemHolds(proj, val) ==> 
                      !SystemHolds(sys, val)
ensures (forall val :: !SystemHolds(proj, val)) ==> 
        (forall val :: !SystemHolds(sys, val))
\end{lstlisting}
\paragraph{}This guarantees that if the projection has no solutions, then the original is unsatisfiable by any valuation. The opposite dependence is not true: if the projection has a solution, it does not guarantee that the original also has a solution. The \emph{Real Shadow} is therefore just an over-approximation used to detect unsatisfiability, but not to establish satisfiability. 

By just eliminating one variable, the algorithm cannot search for solutions. Therefore, the eliminating step has to be performed until no variables are left. The method \texttt{RealShadow} receives an order of elimination and calls \texttt{CombineIneqs} for each variable. After each projection, the system is normalized again, because combining inequalities can lead to duplicate terms or null coefficients. The invariants in the loop keep track of the fact that the produced systems are valid, normalized, the already processed variables do not appear anymore in the systems and the implication that every valuation satisfying the original system also satisfies each successive projection. 
The method returns a variable-free system (property expressed by the predicate \texttt{VariableFreeSystem}) by checking if the terms field in each inequality is empty. In this case, the only remaining inequalities are constants, so the satisfiability is not influenced by a valuation and can be checked by comparing if all inequalities respect the condition: $c_1 \leq c_2$. In Example~\ref{exp:System}, we can see the steps of eliminating variables $x$ and $y$ to obtain a variable-free system.

Unlike the \emph{Real Shadow}, the \emph{Dark Shadow} preserves only those values for which the existence of an integer solution for $x$ can be guaranteed. For a pair of upper bounds and lower bounds, an inequality of the form: $aU - bL > a*b - a - b$ is generated. The condition used to construct the \emph{Dark Shadow} can be derived by contradiction. 

The mathematical argument that justifies the restriction used by the algorithm is formalized in the lemma \texttt{BoundsSatisfyOriginal}. Consider a lower and an upper bound $L \leq ax, bx \leq U, a,b\in \mathbb{Z}_{>0}$. Assume that there is no integer value $x$ that satisfies both limits. In the Dafny proof, we choose \texttt{var i:int := Uv / bv}, other said $i =\left\lfloor \dfrac{U}{b}\right\rfloor$. Since all quantities are integers, the strict inequalities imply the following stronger assertions with a gap of at least \(1\). The same argument is used for \(U < b(i+1)\).
\begin{lstlisting}[frame=tb, belowskip=0pt]
assert av * i + 1 <= Lv;
assert 1 <= Lv - av * i;
assert Uv + 1 <= bv * (i + 1);
assert 1 <= bv * (i + 1) - Uv; // by combining the two gaps, we obtain:
assert av * Uv - bv * Lv <= av * bv - av - bv
\end{lstlisting}
\paragraph{}Thus, the absence of an integer solution implies the inequality above. By contraposition, its negation, $aU-bL>ab-a-b$, guarantees that the interval contains at least one integer value of $x$. This sufficient condition forms the basis of the \emph{Dark Shadow} projection~\cite{dillig_omega}. The existence of the integer witness is obtained from the lemma \texttt{InverseStarStarImpliesIntegerSolution}.

Because all values are integers, a strict inequality can be represented by a non-strict inequality: $a < b <==>  a +1 <= b$. The implementation reuses steps from \emph{Real Shadow}, such as normalization and partition, but imposes supplementary constraints, such as all coefficients and constants must be integers. This property is required by using predicates that check for any Rational $a$ to have $a.den == 1$. This is necessary because the proof for the \emph{Dark Shadow} relies on the fact that the expressions $aU, bL$ and $ab-a-b$ have integer values. 

An important difficulty for verification is that the projected inequality does not contain by itself enough information to reconstruct the value of the eliminated variable. For each generated constraint, the proof has to know the inferior and superior limits from which the new inequality derives. This issue is solved by the \texttt{DarkTraceSystem} structure. Each element \texttt{PairDark} has 3 \texttt{Inequality} fields (\texttt{a, b, c}) that store the two source inequalities and the third generated one. The method that builds this new system, \texttt{PartitionDark}, is crucial to the verification of the algorithm and  has strong specifications to ensure that the pair is in the partitioned system or that the eliminated variable does not appear anywhere in the system anymore. The predicate \texttt{PairDarkHasWitnessForX} states that the pair keeps the necessary algebraic form to reconstruct a value for $x$, and the predicate \texttt{PairDarkIsStarForXRewritten} expresses that the generated inequality is exactly $aU-bL > ab-a-b$.
\begin{lstlisting}[frame=tb, belowskip=0pt]
ghost predicate PairDarkHasWitnessForX(p: PairDark<Inequality, 
                            Inequality, Inequality>, x: string)
{   
    // predicates that state validity ... &&         
    forall val :: EvalInequality(p.c, val) ==> exists n:int ::
            EvalInequality(p.a, UpdateValuation(val, x, n)) &&
            EvalInequality(p.b, UpdateValuation(val, x, n))
}            
ghost predicate PairDarkIsStarForXRewritten(p: PairDark<Inequality, 
                                Inequality, Inequality>, x: string)
    requires JustValidInequality(p.c)
{    
    forall val :: EvalInequality(p.c, val) <==> (ValidTerms(p.c.left.terms) && 
            // other predicates stating validity
            InverseInequalityStarStar(p.b.left, p.a.right, 
            p.b.right.terms[0].coeff,  p.a.left.terms[0].coeff, val))
}
\end{lstlisting}
\paragraph{}A second category of post-conditions describes the completeness of the trace. For each inferior limit from \texttt{ps.neg} and each superior one in \texttt{ps.pos}, the method guarantees the existence of an object \texttt{PairDark}, corresponding in \texttt{trace.pairs}:
\begin{lstlisting}[frame=tb, belowskip=0pt] 
ensures forall j, i | 0 <= j < |ps.neg| && 0 <= i < |ps.pos| :: 
        exists p {:trigger trig(i, j, p)} :: p in trace.pairs && 
                             p.b == ps.neg[j] && p.a == ps.pos[i]
ensures forall p {:trigger p in trace.pairs} :: p in trace.pairs 
                        ==> PairDarkTracksSplit(p, ps) 
\end{lstlisting}
\paragraph{}Together, those two post-conditions show that \texttt{trace.pairs} is exactly the set of dark constraints generated by the pairs \texttt{ps.neg} and \texttt{ps.pos}.

In the implementation, those properties are maintained by the invariants in the loops that build the Cartesian product. The exterior loop fixes an inferior limit, and the interior one covers the superior ones. The method uses an auxiliary variable \verb|pairIdx| to store the position in \texttt{trace.pairs} where the new pair was added. This way, after constructing a new object \texttt{PairDark}, the proof can refer directly to the newly inserted element\\ \verb|trace.pairs[pairIdx]| and can immediately verify its properties: that it was generated from the current limits, that it has the correct dark form, and that it is associated to the pair of indexes $(j,i)$.

The role of \texttt{pairIdx} is important because the final post-condition does not only mention that there exist some dark constraints in the result, but that for each pair of indexes $0 \leq j < |ps.neg|$ and $0 \leq i < |ps.pos|$ there \emph{exists} a corresponding element in \texttt{trace.pairs}. The \texttt{:trigger} notations are necessary for instantiating the quantification used by the SMT solver in Dafny. The expression \texttt{trig(i,j,p)} is used because the witness \texttt{p} needs to be correlated with two indices of the Cartesian product, making it difficult for the solver to verify the condition. Similarly, \texttt{p in trace.pairs} from the second post-condition helps the solver to instantiate the property \texttt{PairDarkTracksSplit} when it encounters an element of a sequence of pairs.

The final property of this projection is \texttt{ensures (exists v :: SystemHolds(outSys, v)) ==> } \texttt{(exists w :: SystemHolds(sys, w))}. Because unsatisfiability is not proven by the \emph{Dark Shadow}, after each variable elimination, we need to search in the \emph{Grey Shadow} branches, which check for a solution in the space between the \emph{Real Shadow} and the \emph{Dark Shadow}.

After equality and divisibility constraints involving the current variable have been handled, Omega Test enters the shadow-based part of the algorithm. It first constructs the \emph{Dark Shadow}. If the recursive call on the dark projection returns \texttt{true}, then the trace information is used to lift a solution of the projection back to a solution of the original system. If the \emph{Dark Shadow} is unsatisfiable, the method constructs the \emph{Real Shadow}. If the real projection is also unsatisfiable, then the original system is unsatisfiable. Otherwise, the method must explore the \emph{Grey Shadow} branches.

The interesting case is when the \emph{Dark Shadow} is unsatisfiable and the \emph{Real Shadow} is satisfiable. In this situation, the implementation follows Pugh's Omega Test~\cite{pugh_omega} by checking the intermediate cases between the two projections, represented as \emph{Grey Shadow} systems. Instead of trying all possible values of the removed variable, the implementation constructs a finite number of subproblems by adding an equality that fixes the position of $ax$ relative to a lower bound: $ax = L+i$. For each lower bound $L\leq ax$, the offset $i$ is traversed in a finite interval, which is calculated from the bound coefficients ($0 \leq i \leq \dfrac{ab-a-b}{b}$). Each choice for $i$ defines a separate \emph{grey branch}, which is checked by a recursive call to the \texttt{OmegaTest} method.

The reason behind this is that we construct the \emph{Grey Shadows} when the \emph{Dark Shadow} is unsatisfiable and the \emph{Real Shadow} is, meaning the system has these properties:

\begin{itemize}
    \item for all pairs $L \leq ax$ and $bx \leq U$: $bL \leq abx \leq aU$ (from the \emph{Real Shadow});
    \item there exists one pair such that $aU \leq bL + ab - a - b$ (from the \emph{Dark Shadow}).
\end{itemize}

Therefore, the new property of those pairs is that $bL \leq abx \leq ab + bL - a - b$ and, if we divide by $b$ (which is non-zero), we obtain $L \leq ax \leq L + \dfrac{ab-a-b}{b}$. Since $x$ is an integer, the possible values of $ax$ in this interval can be checked by considering equations of the form $ax = L+i$, where $0 \leq i \leq \dfrac{ab-a-b}{b}$.

Each \emph{grey branch} adds the equality $ax=L+i$ as a new constraint for the currently tested system. If any of those systems is proved to be satisfiable, then the original is satisfiable too. For example, if one of the lower bounds is $2y+2 \leq 2x$, then $L=2y+2$ and $a=2$. A grey branch can add an equality of the form $2x = 2y+2+i$. For the offset $i=0$, this becomes $2x = 2y+2$.

Before the recursive call, the grey branch is simplified by a GCD check. If the equality is incompatible, the branch is skipped and marked unsatisfiable. If simplification produces a unit coefficient, the equality is substituted and the recursive call proceeds with the current variable removed from the elimination order. Otherwise the remaining non-unit equality is represented through a divisibility constraint added to the system.

For example, the grey equality $4x = 2y + 6$ can be simplified by the GCD $2$, obtaining $2x = y + 3$. On the other hand, if the branch produces $4x = 2y + 5$, then the same GCD does not divide the constant term, so the branch is incompatible and can be marked unsatisfiable without a recursive call.

Divisibility constraints are handled by splitting the problem into finitely many residue cases. If the current variable $x$ occurs in a divisibility constraint, \texttt{OmegaTest} introduces a fresh quotient variable $q$ and considers residues modulo the product of the divisors that appear in the system. For each residue $r$, the method replaces $x$ by an expression of the form $mq + r$, where $m$ is this product. This removes $x$ from the divisibility constraints and the recursive call continues with the fresh variable placed at the beginning of the elimination order. 

For example, a divisibility constraint $3 \mid 2x + y$ is handled by considering the possible residues of $x$ modulo $3$: $x = 3q, x = 3q + 1, x = 3q+2$. Each branch removes $x$ from the divisibility constraint and introduces the fresh quotient variable $q$. If several divisibility constraints are present, the implementation uses the product of their divisors as a common modulus.

This does not test all possible values of $x$, but only the values that can occur close to the lower bound, in the interval where the \emph{Dark Shadow} may have lost possible integer solutions.

At the end of the exploration of the \emph{grey branches}, if no subproblem is satisfiable, then the method proves the unsatisfiability of the original system by contradiction. We assume there exists a valuation that satisfies the original system. Because the \emph{Dark Shadow} is already proven unsatisfiable on this path, we use a lemma to show that this valuation has to satisfy at least one of the grey equalities. But since all grey branches were already marked as unsatisfiable by \texttt{GreyBranchUnsat}, this contradicts the existence of the valuation. The final argument of the method \texttt{OmegaTest} is
\begin{lstlisting}[frame=tb, belowskip=0pt] 
ensures sat ==> (exists v :: SystemHolds(sys, v))
ensures !sat ==> (forall v :: !SystemHolds(sys, v))
\end{lstlisting}

\section{Termination  of the recursive method}\label{sec:termination}
\paragraph{}Proving the termination of the algorithm is not trivial, because a direct implementation of the idea behind the Omega Test leads to an infinite loop, illustrated below:

\[
\begin{cases}
    - 2x + y \leq -1 \\
    2x - y \leq 1
\end{cases}
\xrightarrow{\text{Partition}}
\begin{cases}
    y+1 \leq 2x \\
    2x \leq y+1.
\end{cases}
\]
\paragraph{}Suppose the first variable considered for elimination is $x$. The algorithm will process the inequalities such that the coefficient of $x$ is always positive. In this case, we obtain a lower and an upper bound for $x$: $L = y + 1$, $U = y + 1$ and $a = b = 2$. The Real Shadow condition $bL 
\leq aU$ is satisfied, because $2(y+1) \leq 2(y+1)$. However, the Dark Shadow condition, $aU-bL > ab - a- b$ becomes $0 > 0$, which is false. Therefore, the algorithm enters the Grey Shadow case.

The only grey offset is $i = 0$ (because $0 \leq i \leq \dfrac{2*2-2-2}{2}$), so the grey branch adds the equality $2x = y + 1$. If the implementation does not treat this equality using divisibility, it may represent it again as two inequalities $y+1 \leq 2x \qquad\text{and}\qquad 2x \leq y+1$. Therefore, the recursive call receives essentially the same system as before, still containing the variable $x$. The next Grey Shadow step can add the same equality again, producing the sequence $S, \quad S \cup \{2x = y + 1\}, \quad S \cup \{ 2x=y+1, 2x = y+1\}, \quad \ldots$, so instead of decreasing the recursive problem, the system keeps extending with repeated constraints.

Therefore, practical implementations rely on divisibility constraints. The equality $2x = y + 1$ is solvable over integers when $2 \mid y +1$. After recording this divisibility condition, the method can split the problem into finitely many residue cases instead of repeatedly adding the same grey equality.

To our knowledge, we are the first to formalize a verified termination argument for the Omega Test.

We justify the termination of the recursive method \texttt{OmegaTest} by using the lexicographic measure \[ \begin{array}{l} \qquad (|\texttt{elimOrder}|,\\ \qquad \texttt{IsFirstVarInDiv(sys, elimOrder)},\\ \qquad |\texttt{sys.equalities}|,\\ \qquad \texttt{NumberVarSideConstraints(sys, elimOrder)}).\end{array} \]

The first component decreases when a variable is eliminated by substitution or by a shadow projection and the recursive call is made with \texttt{elimOrder[1..]}. 

The second component handles the case where the current variable occurs in divisibility constraints: these constraints are expanded into finitely many residue branches using a fresh quotient variable, after which the head variable is no longer in a divisibility constraint. The ghost function \texttt{IsFirstVarInDiv(} \text{sys, elimOrder)} is equal to $1$ when the first variable in the elimination order occurs in some divisibility constraint, and $0$ otherwise. Before residue splitting, $x$ occurs in a divisiblity constraint, so this component is $1$. After replacing $x$ by $mq + r$, the fresh variable $q$ is introduced only through the residue equality and does not occur in the remaining divisibility constraints. Therefore, in the recursive call, \texttt{IsFirstVarInDiv(sys, elimOrder)} becomes $0$ and the lexicographic measure decreases even though the length of the elimination order remains unchanged.

The third component, \texttt{sys.equalities}, handles recursive calls that keep the same elimination order but remove a trivial equality, such as $0 =0$. The fourth component, \texttt{NumberVarSideConstraints(sys, elimOrder)}, counts the equalities and divisibility constraints that still mention the current variable. It decreases in simplification steps where the method rewrites an equality so that the current variable no longer occurs in it, while keeping the same elimination order.

The residue split itself is also finite. The loop over residues uses the measure $m-r$, where $m$ is the product of the divisors and $r$ is the current residue. Since all divisors are positive, $m>0$, the loop checks only the residues $0, \ldots, m-1$. Each recursive call made inside this loop has a strictly smaller lexicographic measure, so introducing the fresh quotient variable does not create an infinite recursion.

\section{Limitations and Future Work}\label{sec:Limitations}
\paragraph{}Although this implementation verifies the correctness properties of the elimination procedure, it is not intended to be a performance-oriented solver. The focus of the work is to connect the executable Dafny implementation with the mathematical correctness proof of the Omega Test. As a consequence, the implementation uses verification-friendly representations, such as immutable sequences and explicit rational numbers, which make the proofs easier to state but introduce overhead through repeated traversals, copying, normalization, and reconstruction of expressions and systems.

Verification performance is also a limitation. On our machine (13th Gen Intel Core i5-1334U, 15.7 GB RAM, 512 GB NVMe SSD), verifying the full Dafny development takes approximately 49 minutes and 22 seconds. This reflects the size of the proof obligations generated by the Real, Dark, Grey, equality, and divisibility cases, together with the auxiliary lemmas and loop invariants required to prove correctness and termination. Future work could improve proof performance by reducing repeated assertions, moving common proof patterns into smaller lemmas, and using verification-splitting attributes only where they are necessary.

The elimination procedure follows an elimination order generated by a method that just finds all variables in a system one by one. Because different elimination orders can lead to a smaller number of generated inequalities or grey branches, a future extension could implement and verify heuristics for selecting the next variable to be eliminated.

Finally, the current implementation focuses mainly on deciding satisfiability. Another useful extension would be explicit model reconstruction, so that the algorithm would not only return whether a system is satisfiable, but also produce a valuation when one exists. Some parts of the proof already keep enough information to justify how a solution of a projection can be lifted back to a solution of the original system, especially in the Dark Shadow phase.

\bibliographystyle{eptcs}
\bibliography{generic}

\end{document}